# Google Scholar and the gray literature:
## A reply to Bonato's review♣

**Enrique Orduna-Malea[1], Alberto Martín-Martín[2] &**
**Emilio Delgado López-Cózar[2]**

[2] *EC3: Evaluación de la Ciencia y de la Comunicación Científica, Universitat Politècnica de València (Spain)*
[1] *EC3: Evaluación de la Ciencia y de la Comunicación Científica, Universidad de Granada (Spain)*

## ABSTRACT

Recently, a review concluded that Google Scholar (GS) is not a suitable source of information "for identifying recent conference papers or other gray literature publications". The goal of this letter is to demonstrate that GS can be an effective tool to search and find gray literature, as long as appropriate search strategies are used. To do this, we took as examples the same two case studies used by the original review, describing first how GS processes original's search strategies, then proposing alternative search strategies, and finally generalizing each case study to compose a general search procedure aimed at finding gray literature in Google Scholar for two wide selected case studies: a) all contributions belonging to a congress (the ASCO Annual Meeting); and b) indexed guidelines as well as gray literature within medical institutions (National Institutes of Health) and governmental agencies (U.S. Department of Health & Human Services). The results confirm that original search strategies were undertrained offering misleading results and erroneous conclusions. Google Scholar lacks many of the advanced search features available in other bibliographic databases (such as Pubmed), however, it is one thing to have a friendly search experience, and quite another to find gray literature. We finally conclude that Google Scholar is a powerful tool for searching gray literature, as long as the users are familiar with all the possibilities it offers as a search engine. Poorly formulated searches will undoubtedly return misleading results.

## KEYWORDS

**Google Scholar; Gray literature; Conference proceedings; Guidelines; Academic search engines**

**Corresponding author**
**Emilio Delgado López-Cózar.** edelgado@ugr.es

♣ This work has been rejected in the Journal of the Medical Library Association (JMLA), both the full version (24th December, 2016) and a letter to editor version (31st January, 2017).

## 1. INTRODUCTION

Google Scholar, currently the preferential scientific literature's search tool platform to find scientific literature across all disciplines (Orduna-Malea et al, 2016), particularly in the Biomedicine (Vine, 2006; Hasse et al, 2007; Freeman et al, 2009; Kulkami et al, 2009; Mastrangelo et al, 2010; Tober, 2011; Nourbakhsh et al, 2012; Bramer, 2016; Trapp, 2016), is now also being used as a source of data to elaborate systematic reviews (Boeker, Vach & Motschall, 2013; Gehanno, Rollin & Darmoni, 2013; Giustini & Boulos, 2013; Bramer, Giustini & Kramer, 2013; 2016).

Recently, a review (Bonato, 2016) inquired into the search options in Google Scholar and Scopus for finding gray literature concluding – after running few queries – that Google Scholar is not such an ideal information source "for identifying recent conference papers or other gray literature publications", leaving much to be desired as a source for scientific information.

The goal of this working paper is to prove how fallacious and disproportionate this statement is, since its conclusions are derived from few specific searches and sustained on erroneous search strategies. The secondary goal of this working paper is to demonstrate that Google Scholar can be an effective tool to search and find gray literature, as long as appropriate search strategies are used

## 2. METHOD

We will take as a sample the same two case studies performed by Bonato. The first aims to locate conference papers on the drug panobinostat (using three different search variations) whereas the second aims to locate one specific document, the "Guidelines for the Use of Antiretroviral Agents in HIV-1-Infected Adults and Adolescents" (using one search variation), published by the US Department of Health and Human Services.

For each case, first we highlight how Google Scholar operates the search strategies performed by Bonato. Second, we propose an expert search to properly solve each case study. And third, we generalize each case study to compose a general search procedure aimed at finding gray literature in Google Scholar for two wide selected case studies.

As regards the first case study, we propose to find all contributions belonging to a congress (the ASCO Annual Meeting). As with the second case study, we propose a procedure to find indexed guidelines as well as gray literature within medical institutions (National Institutes of Health) and governmental agencies (U.S. Department of Health & Human Services).

To do this we use the Google Scholar's advance search as well as the following available search filters and commands:

- ***"allintitle" command***: restricts results to those containing all the query terms you specify in the title of the document
- ***"site" command***: get results from certain sites or domains
- ***"author" command***: restrict results to documents signed by specified authors
- ***Publication field***: restrict results to specific sources
- ***Date of publication field***: restrict results to specific year of publication
- ***Quotation marks***: the results only include documents containing the same words in the same order as the ones inside the quotes
- ***Patents and citations***: the user may select either the inclusion or exclusion of citations and patents in the search engine results page

By way of illustration, the specific use of these commands will be detailed when describing each proposed procedure in the results section. All searches were performed on August 2016 in the Google Scholar's English version.

# 3. RESULTS

## Case study A: Conference papers

***Bonato's search. Variation-1: Panobinostat conference paper***

This search may provoke the following shortcomings:

- Google Scholar will retrieve all documents (including cited references) containing all these three keywords simultaneously.
- Each keyword may be located elsewhere (title, abstract, body, bibliographic references, etc.) as they will be treated independently.

The noise of this query is therefore terrific. Bonato obtained a total of 5480 results.

***Bonato's search. Variation-2: Panobinostat ASCO Annual Meeting***

The name of the congress in which a contribution on panobinostat was presented is now introduced. However, the same previous limitations still operate. The number of results drops from 5480 to 848 simply because the four terms used are more restrictive.

***Bonato's search. Variation-3: Phase I/II study of the combination of panobinostat (PAN) and carfilzomib (CFZ) in patients (pts) with relapsed or relapsed/refractory multiple myeloma (MM)***

The title of the conference paper is now included. This query is designed again upon the incorporation of single independent keywords. Quotation marks to delimit the keywords' order are not even used. Bonato obtains just one result, claiming however that this does not correspond to the original contribution.

### *Search proposal*

We directly query for the title of the contribution within quotation marks in the title search field of the Google Scholar's advance feature, which directly introduces the allintitle command (Figure 1), finding one result which contains five versions (Figure 2), of which two correspond with ASCO 2015 meeting, two other belong to the American Society of Hematology (ASH) congress, and the remaining version belongs to Blood, a journal edited by ASH (Figure 3).

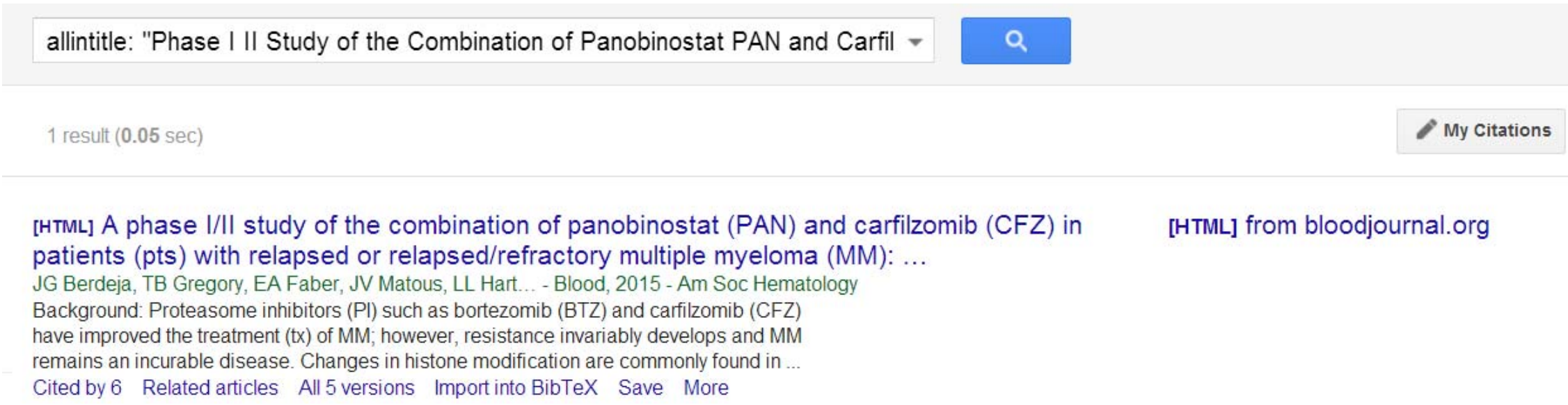

**Figure 1. A query for ASCO Annual Meeting contribution: query**

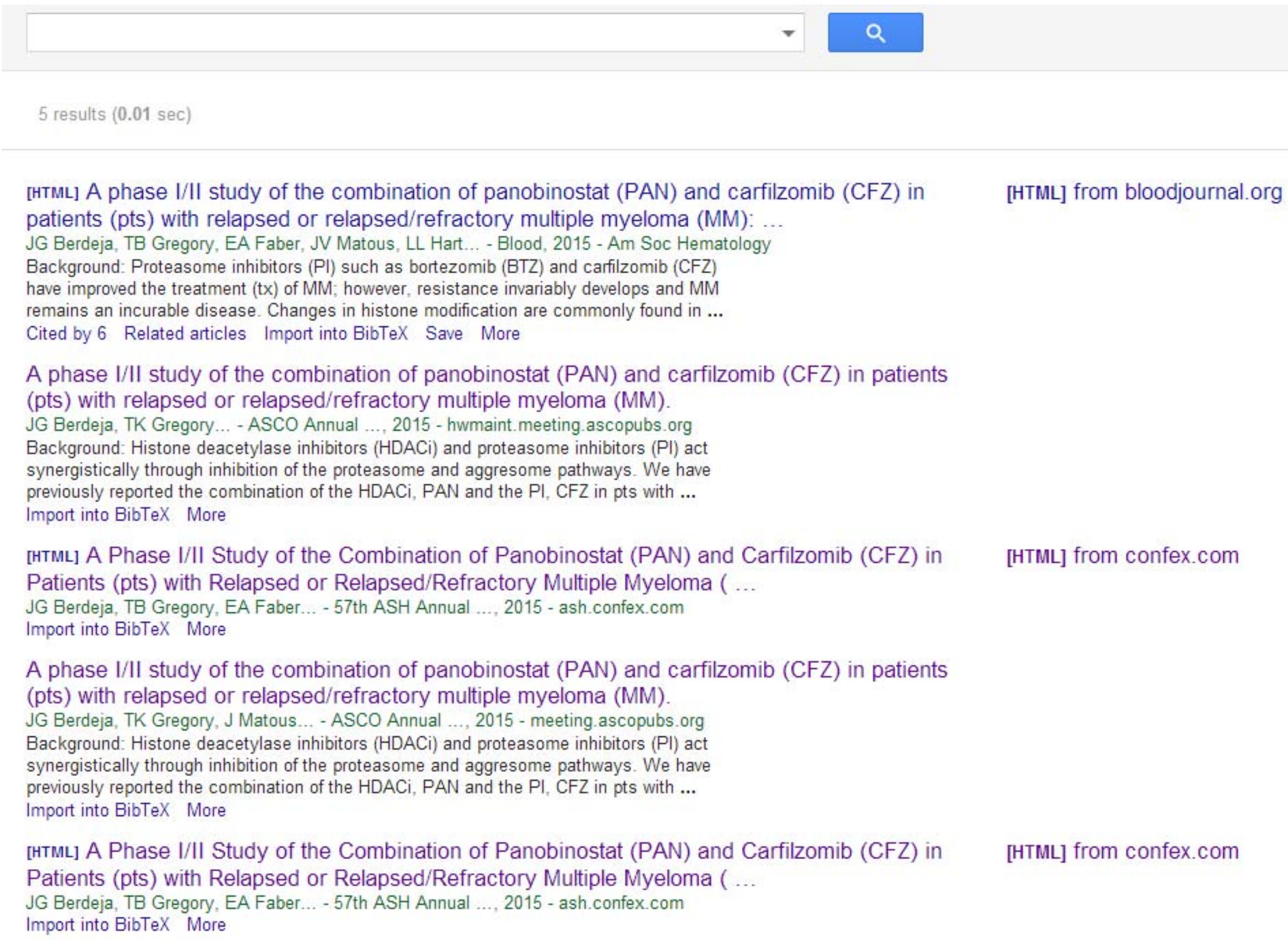

**Figure 2. A query for ASCO Annual Meeting contribution: versions located**

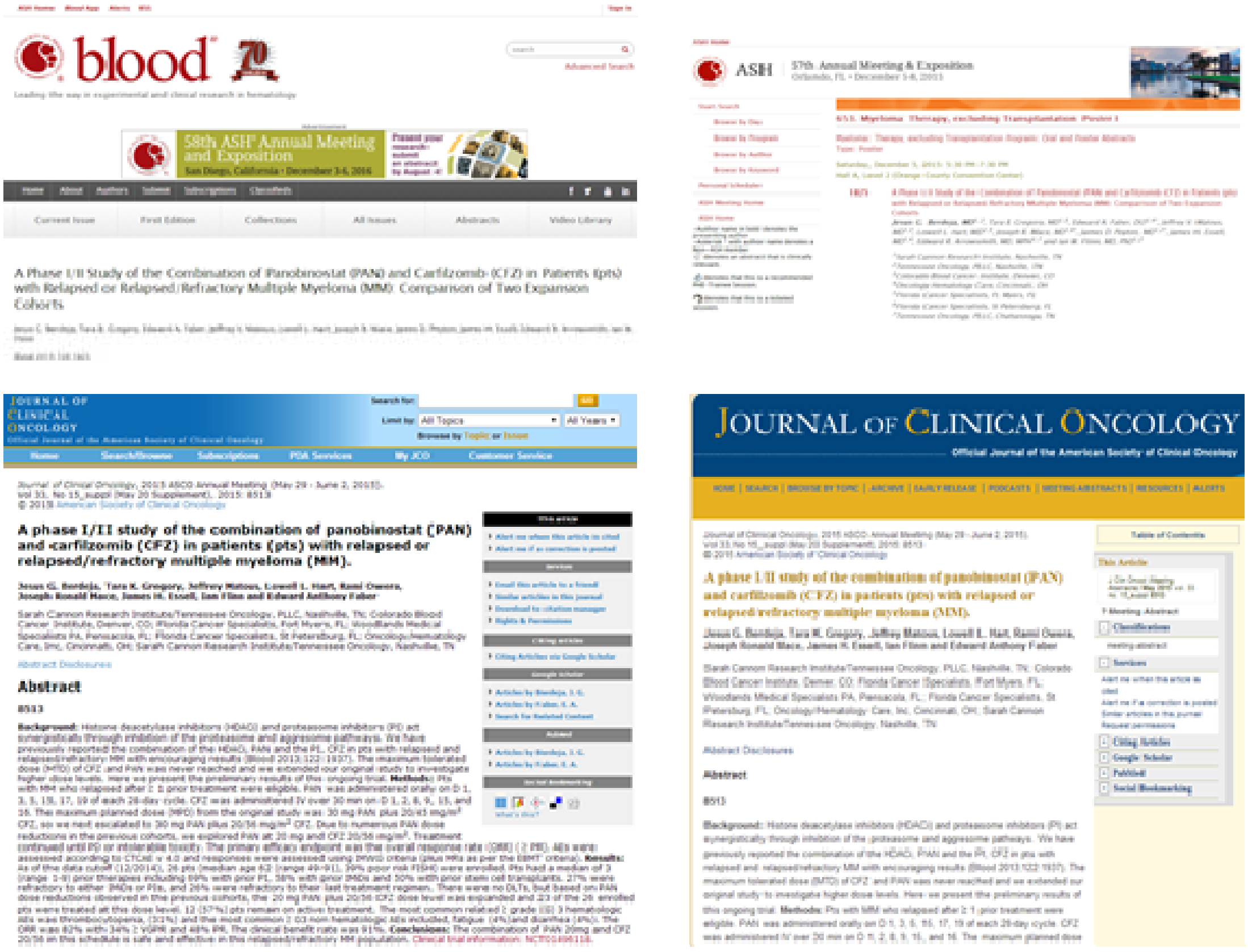

**Figure 3. A query for ASCO Annual Meeting contribution: publications**

## *Searching for a general conference: ASCO Annual Meeting*

A first attempt to search for a conference would be constructing a query based on the name of the congress (ASCO Annual Meeting). By doing this, we find out of 138000 documents (Figure 4). However, this Boolean search with an implicit “AND” is not appropriated since the three keywords may be located in any section of the retrieved documents independently, as previously commented.

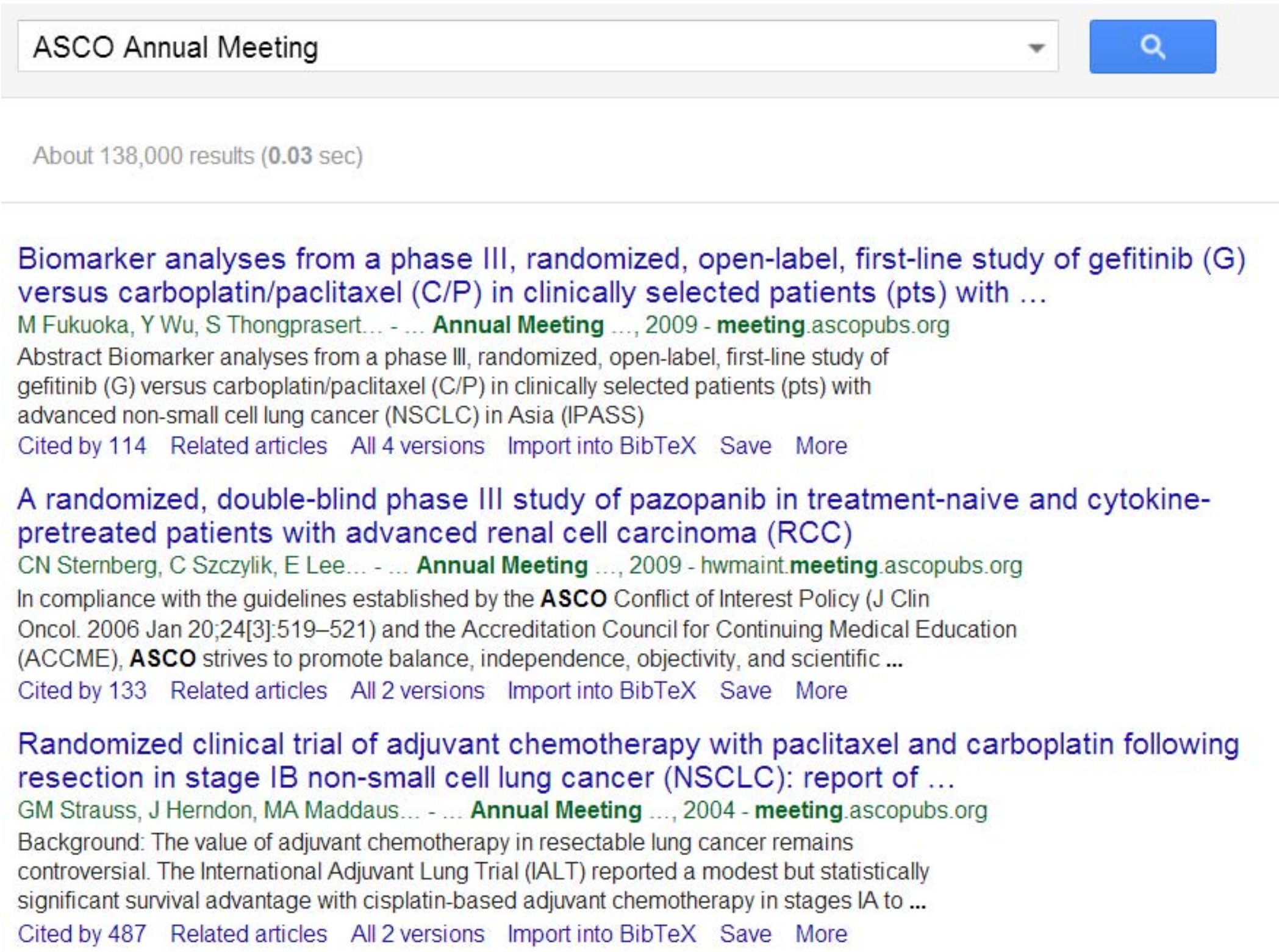

**Figure 4. Search using the Boolean "AND" without quotation marks delimiting character chain**

We can delimit the year of publication to 2015 (Figure 5). In this case, the results reduce to 10400 documents. However, Google Scholar provides not only documents published in the conference but also documents citing those documents.

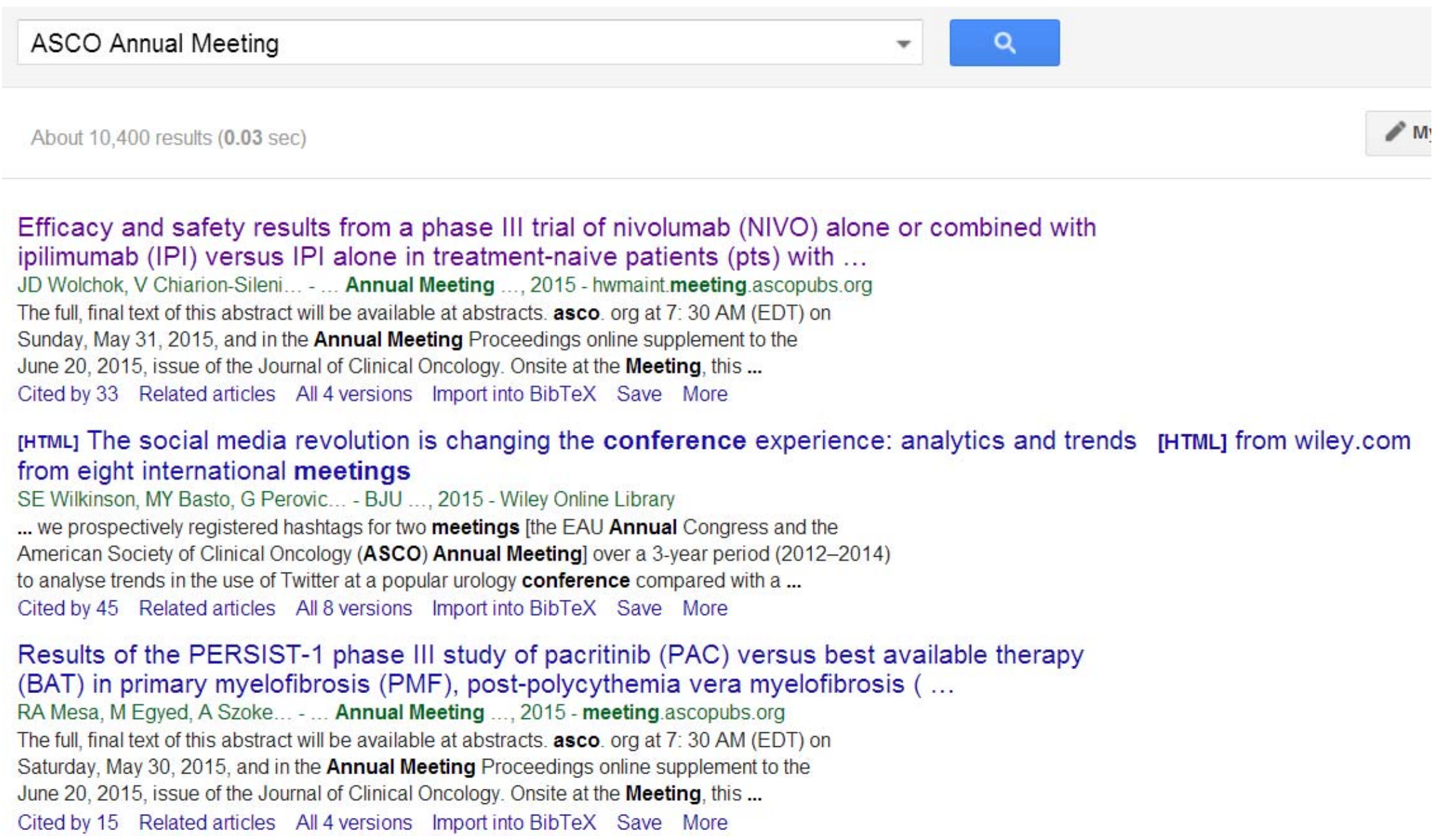

**Figure 5. Search using the Boolean "AND" without quotation marks delimiting character chain, delimiting year of publication to 2015**

We can add a quotation mark to the query in order to delimit the presence of the three keywords exactly in this sequence order, reducing to 7860 documents (Figure 6).

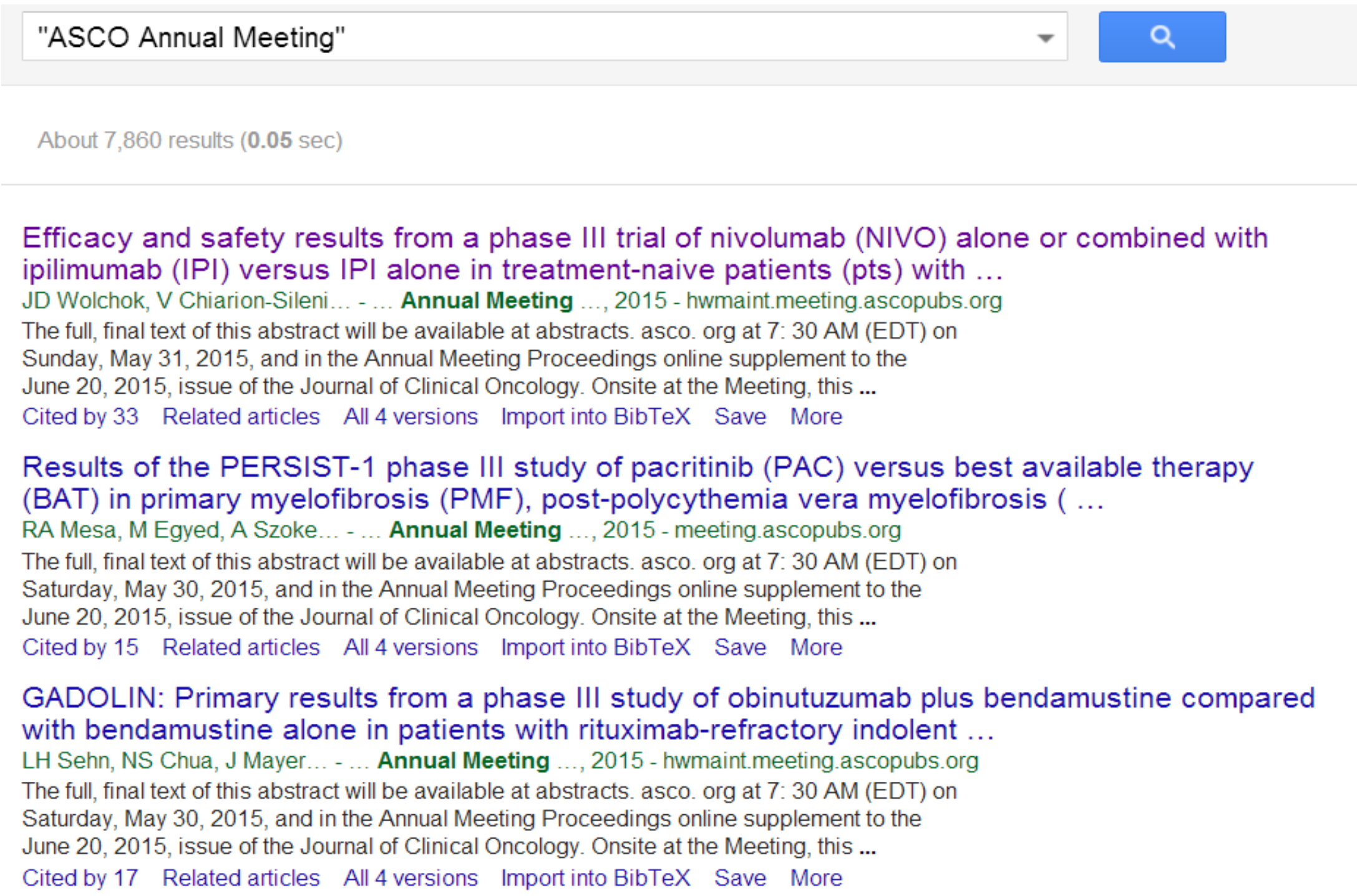

**Figure 6. Search using the Boolean "AND" with quotation marks delimiting character chain, delimiting year of publication to 2015**

Then, we can eliminate the citing documents by adding "ASCO Annual Meeting" in the "Publication" field (Figure 7). Additionally, the option "citations included" should be unchecked, recovering a final set of 6260 results.

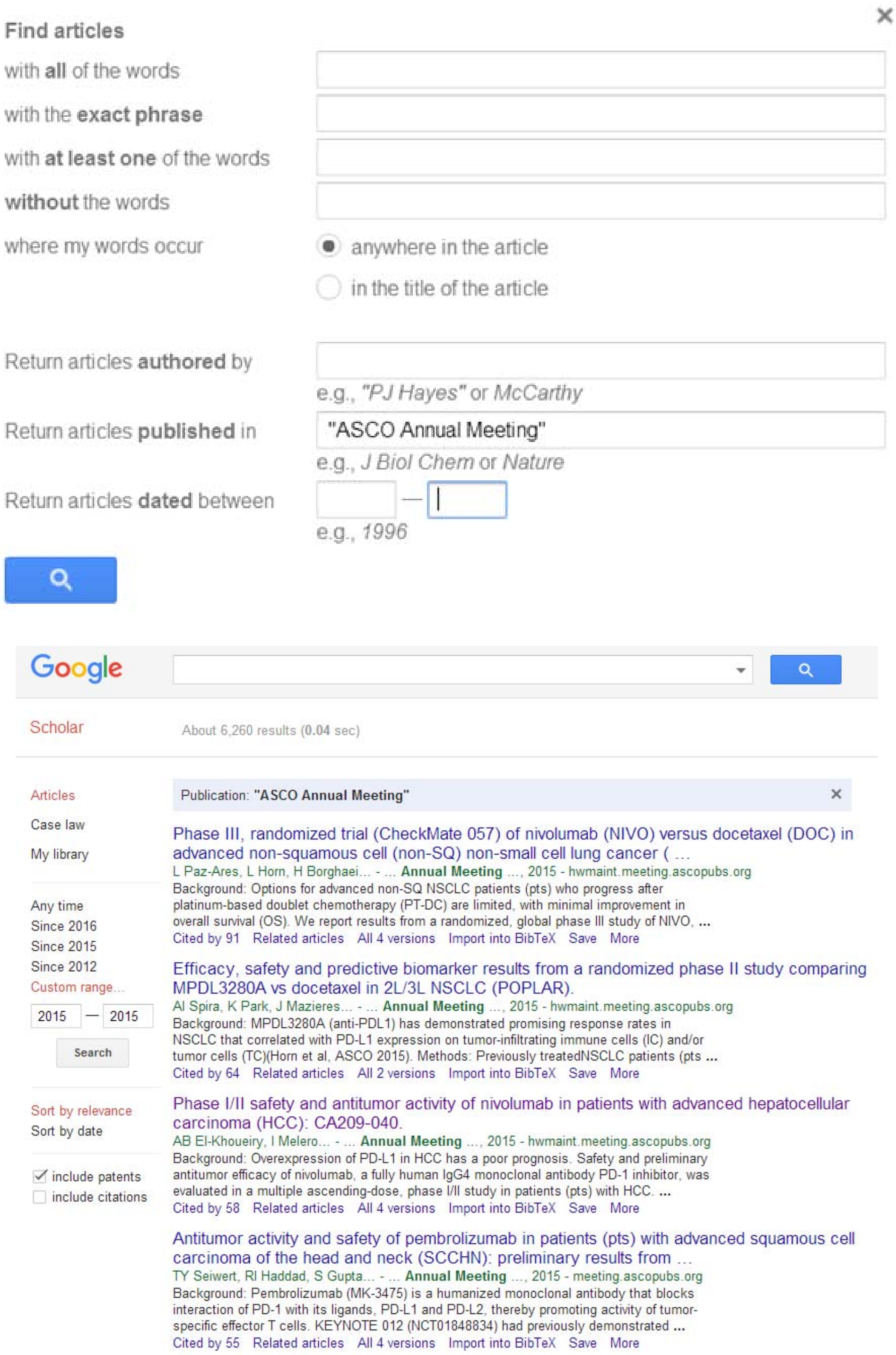

**Figure 7. Google Scholar's advanced search features: selecting publication source (up) and visualizing results (bottom)**

The results page shows URLs where these documents are online available. We can easily verify the existence of the meeting.ascopubs.org and hwmaint.meeting.ascopubs.org web domains (Figure 8). Since the conference proceedings are available in these two different locations, the corresponding records are likely to be duplicated.

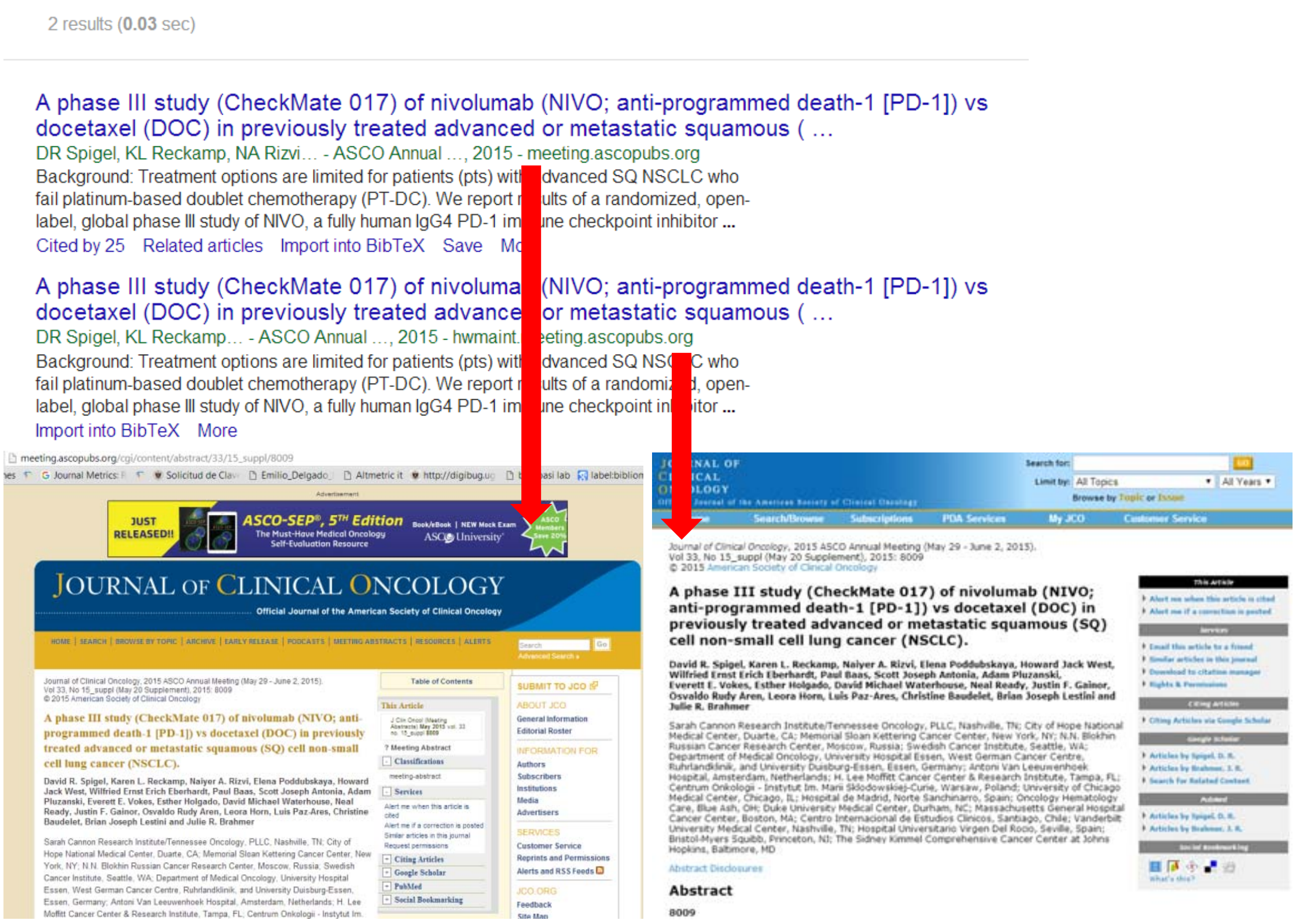

**Figure 8. Different online locations for ASCO Annual Meeting in Google Scholar**

Then we can finally use the "site" search command to restrict results to those documents hosted within one of the two identified web domains, which may provide an accurate value of the Google Scholar's index for this particular conference. The following query site:meeting.ascopubs.org – site:hwmaint.meeting.ascopubs.org provides a final count of 3060 results (Figure 9).

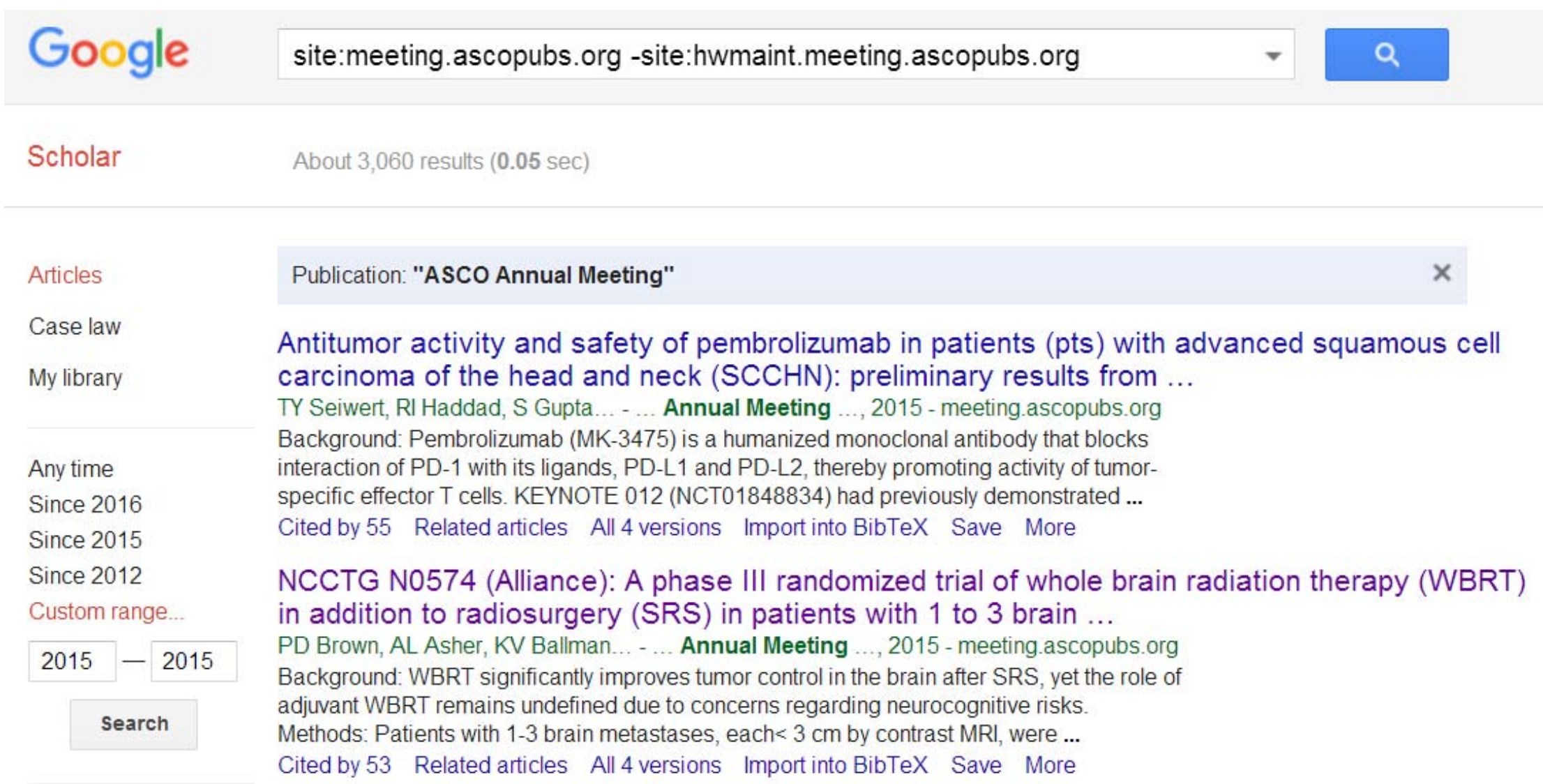

**Figure 9. A query for ASCO Annual Meeting in Google Scholar using "site" search command to specific web domains**

Once located the congress, we can even search easily for documents on specific topics. Coming back to the Bonato's original strategy, we can add a topical keyword to the previous query panobinostat site:meeting.ascopubs.org – site:hwmaint.meeting.ascopubs.org, obtaining the desired result in each of the two web domains previously identified. The following reference URL can be used for re-executed iterations Figure 10):
https://scholar.google.com/scholar?as_ylo=2015&q=panobinostat+site:meeting.ascopubs.org+%E2%80%93site:hwmaint.meeting.ascopubs.org&hl=en&as_sdt=0,5

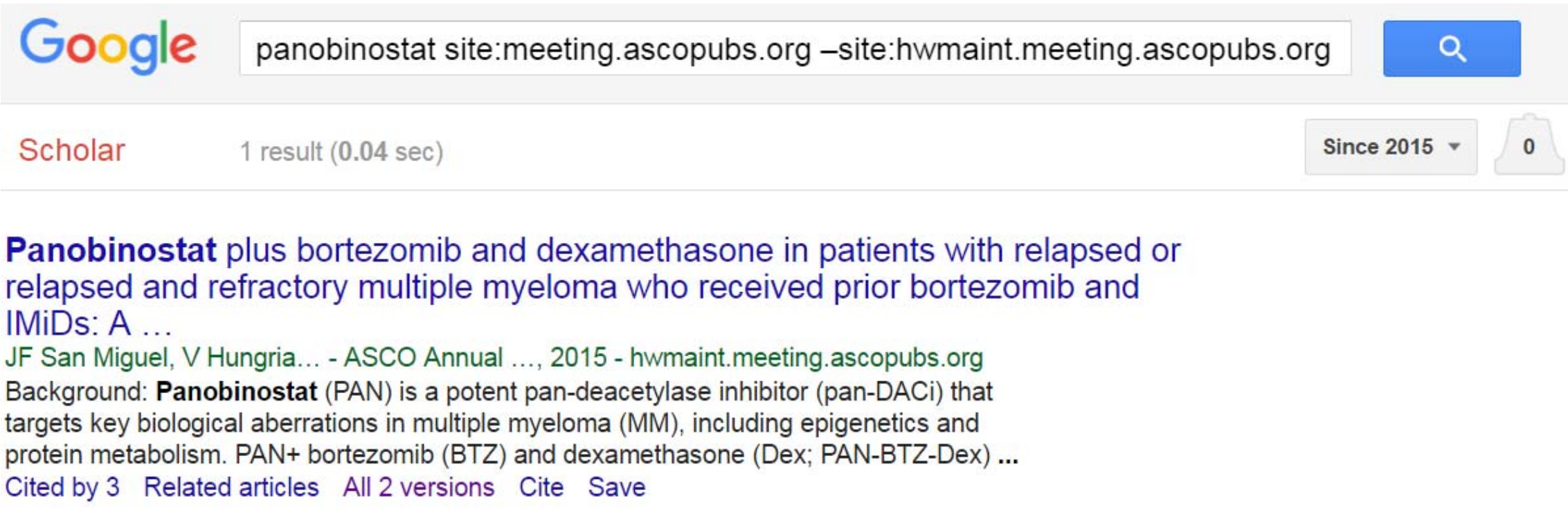

**Figure 10. Complete query for ASCO Annual Meeting contribution adding topic keyword**

## Case study B: searching for guidelines on particular topic

### *Bonato's Search: Guidelines for the Use of Antiretroviral Agents in HIV-1-Infected Adults and Adolescent*

Bonato's search strategy is based on the simple addition of words (a search keyword under Google's jargon) obtaining about 17700 results. However, we should highlight the following concerns:

- The date of publication is not delimited. Filtering documents from 2015 onwards, the number of results drops from 17700 to about 2770.
- Quotations are not used. Google Scholar will check the appearance of all keywords regardless their position and order. This procedure is discouraged when searching for document titles.
- Document source is not delimited.

### *Search proposal*

We directly include the title of the document within quotation marks in the title search field (or directly through the allintitle command) (Figure 11). We find one result that corresponds with the targeted document, which is cited five times: three citations correspond with previous editions (2006 and 2008), one is correct, and the remaining citation is unknown (Figure 12).

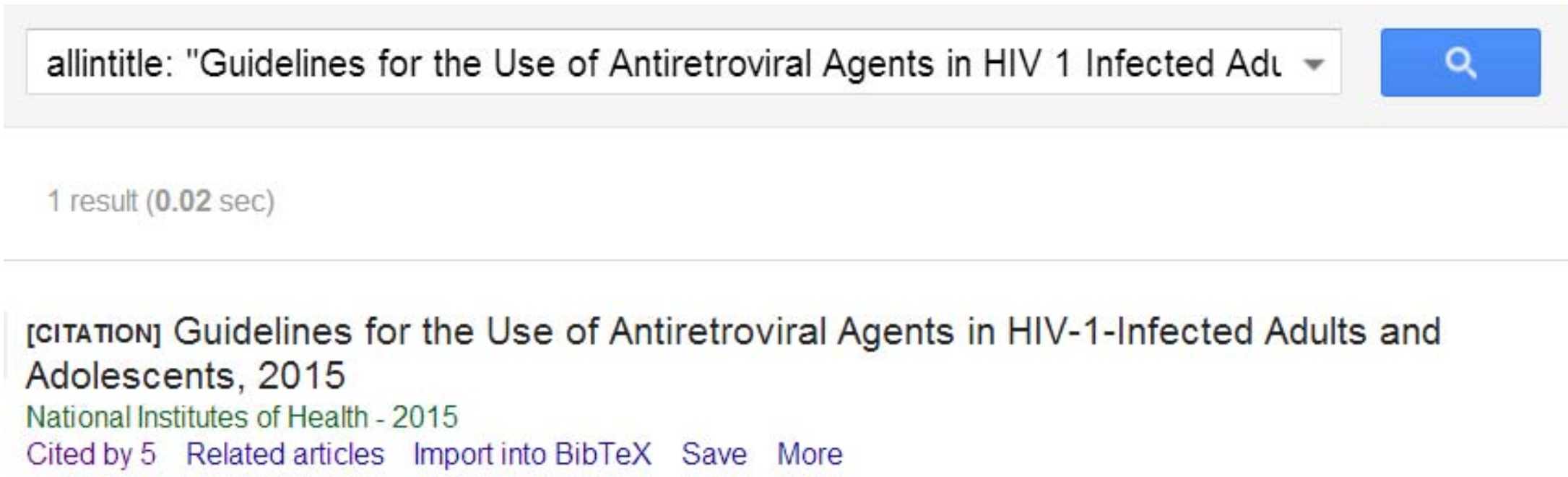

**Figure 11. A query for Guidelines, using the "allintitle" search command**

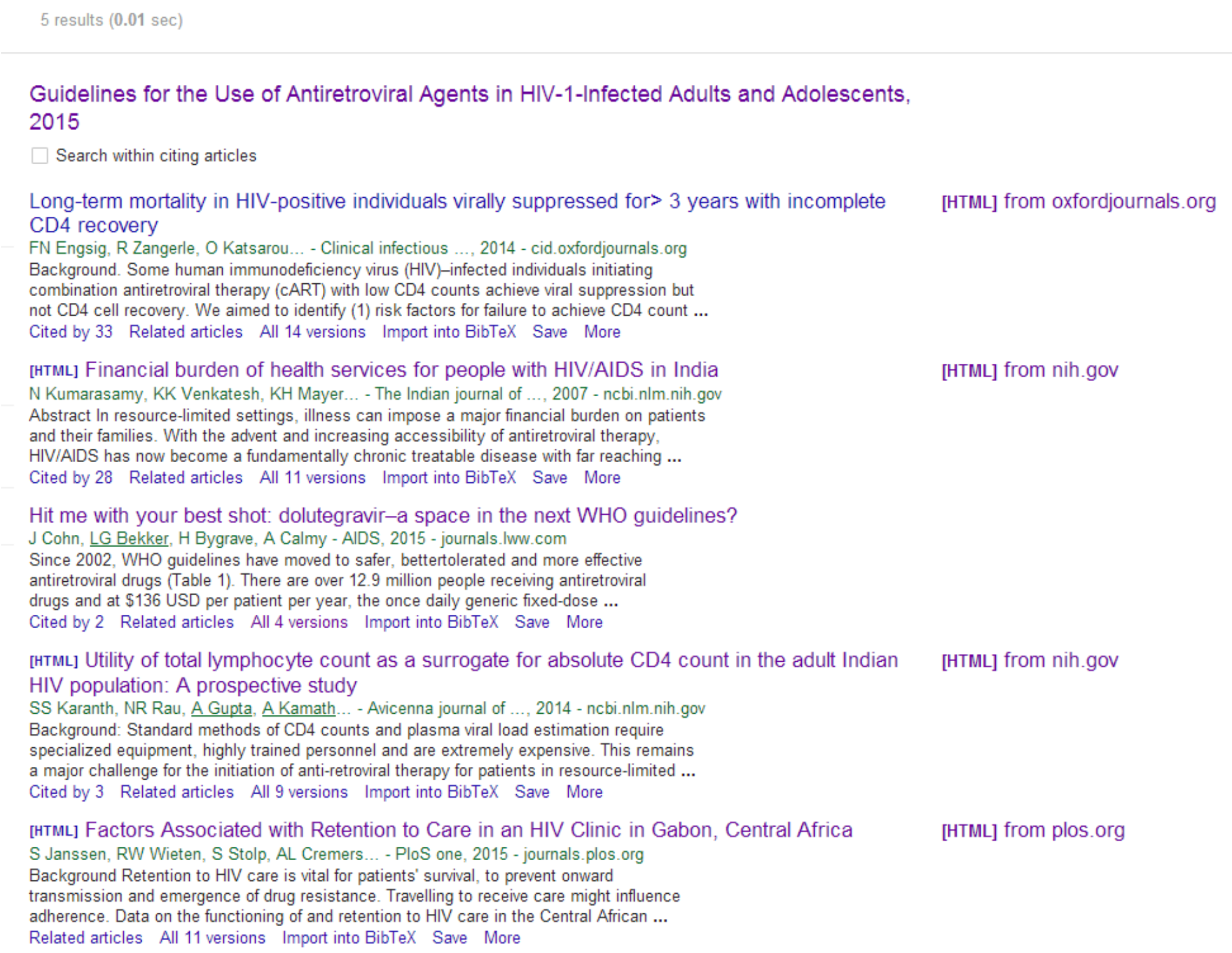

**Figure 12. Citations received by the Guidelines according to Google Scholar**

However, the record is not directly indexed by Google Scholar, appearing only as a citation (with the [C] mark) because it was cited in another document already indexed by Google Scholar. The reason probably relies in the fact that the corresponding PDF file weights approximately 5 MB, being this size the limit to be directly indexed in the database.

### *Searching for guidelines*

We can employ two general approaches to find medical guidelines in Google Scholar. The first one has been already explained in the expert search proposal, and consists of searching directly for a specific title. Of course, it operates only when the user knows such title. A complementary approach consists of searching for guidelines covering specific topics, where the inclusion of search keywords is mandatory. For example: allintitle: "guidelines" heart OR cardiology. This strategy retrieves all those guidelines in which the terms heart or cardiology appear in the title (about 4870 results; if citations are excluded the results drop to 1690). If we perform the equivalent search in Pubmed ((guidelines[Title]) AND (heart[Title]) OR cardiology[Title])), we obtain a total of 1176 results.[1]

One step further consists of specifying all those guidelines published by one specific institution. We can then operate in two different ways. On the one hand can apply a search using the “author” search command: allintitle:"guidelines" author:"National Institutes of Health". This query sheds 373 results (Figure 13). We obtain only three hits when restricting results to 2015, among which we can find Bonato’s targeted guideline.

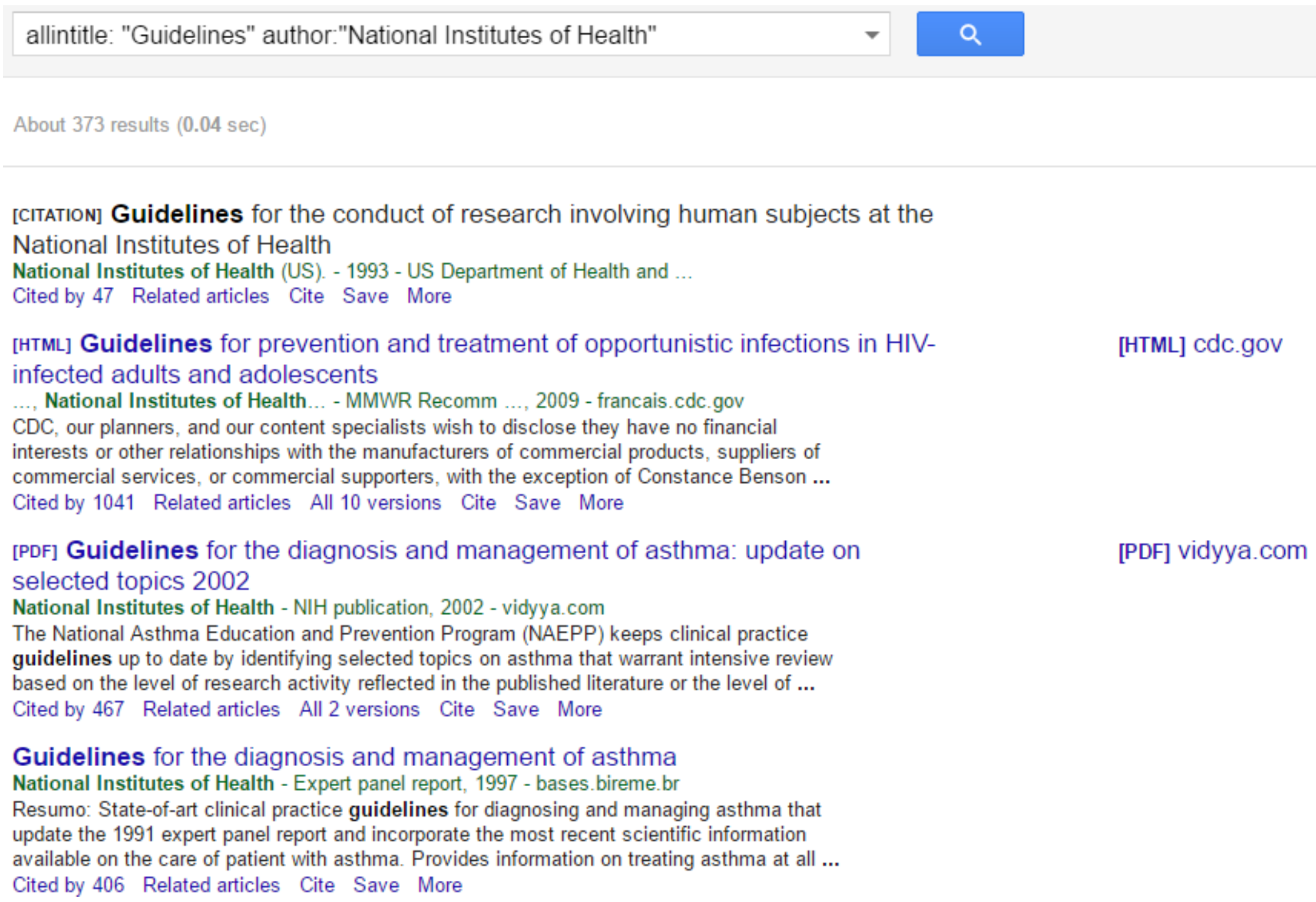

**Figure 13. Use of Allintitle and Author search commands**

On the other hand, we may wish to employ a more generic query where the institution is signaled within quotation marks instead of the author search command. For example: allintitle: guidelines "American Heart Association". This search sheds 566 results, whereas the equivalent search in Pubmed guidelines[Title] AND (American Heart Association[Title]), gives only 26.

### *Searching documents in medical institutions*

A second possible generalization consists of searching for gray literature in medical institutions. We expose two paradigmatic examples: National Institute of Health (NIH), and the U.S. Department of Health and Human Services (HHS).

[1] Savvy users may notice whether parentheses are used to contain a query or forms part of the query itself. Some reviewers unfortunately did not.

a) National Institute of Health (NIH) (http://www.nih.gov)

Since Pubmed repository (ncbi.nlm.nih.gov) is indexed in Google Scholar, any document indexed in this database, whether or not published by NIH, will be retrieved if we do nothing about it. Therefore, Pubmed should be filtered. We can also restrict only those documents in specific file formats (PDF and HTML mainly). For example, the following query (allintitle: guidelines site:nih.gov -site:ncbi.nlm.nih.gov filetype:pdf) gives a total of 9 guidelines (Figure 14). If eliminating the term guideline in the title of documents, we obtain 12000 results.

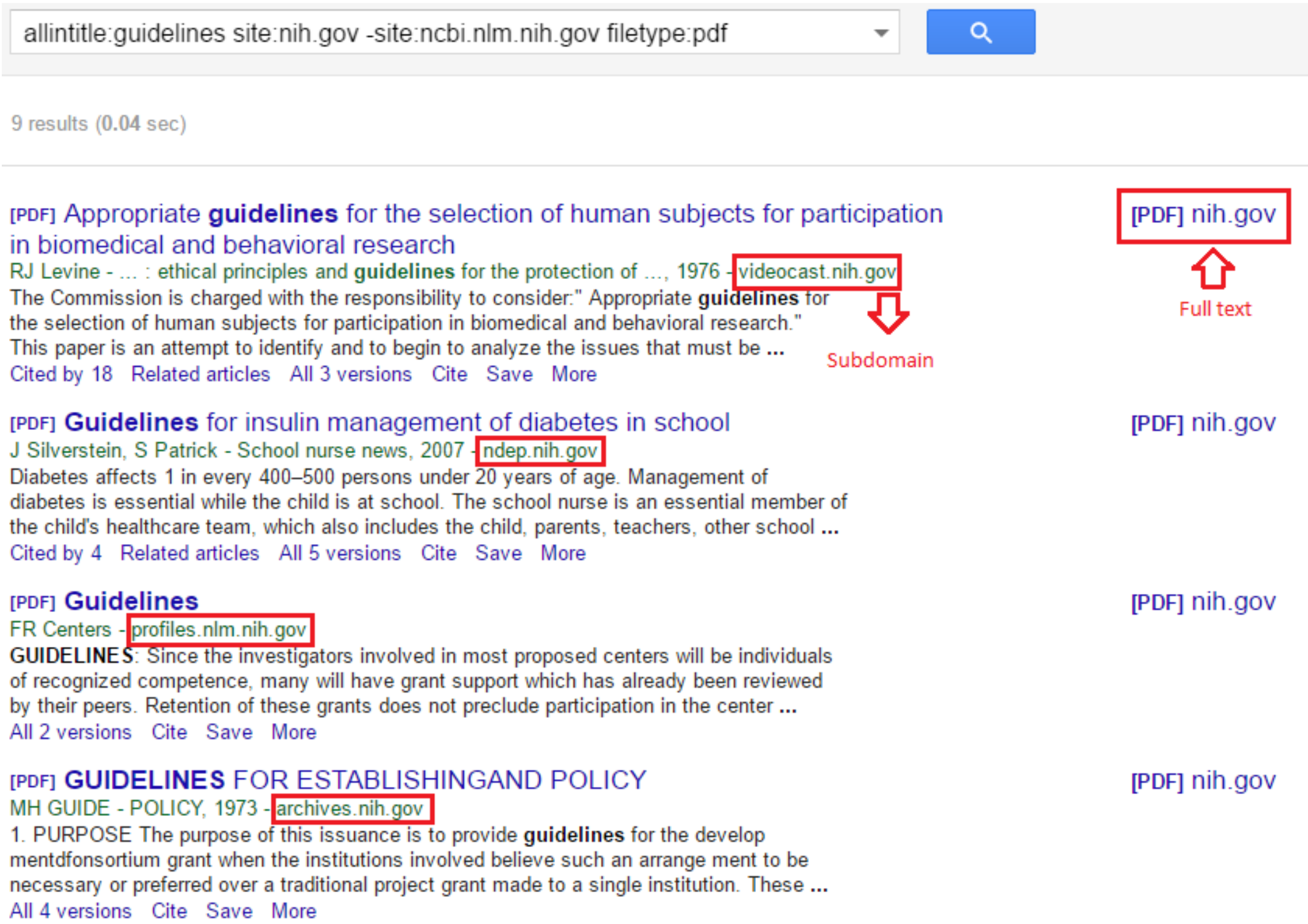

**Figure 14. Guidelines in PDF file format, excluding Pubmed**

We can observe different subdomains within the nih.gov website which are hosting gray literature. A recommended procedure consists of identifying the main subdomains analyzing them individually. In the table 1 we provide an example for the NIH. In this case we have identified (by manually checking) all those centers allocated within the nih.gov web domain. Additionally, we have added the Archived sites services, which stores online historical information.

For each of the websites we can check the total number of documents indexed by Google Scholar (site:*.nih.gov). Excluding Pubmed, the number of documents indexed by Google Scholar is high (15300), showing a skewed distribution among centers.

**Table 1. Documents hosted by National Institutes of Health (NIH) indexed in Google Scholar (as of October 2016)**

| SOURCE | URL | DOCS |
|---|---|---|
| *National Institutes of Health | nih.gov | 15300 |
| *National Library of Medicine | nlm.nih.gov | 9850 |
| National Institute of Environmental Health Sciences | niehs.nih.gov | 2870 |
| National Institute of Mental Health | nimh.nih.gov | 840 |
| National Institute on Alcohol Abuse and Alcoholism | niaaa.nih.gov | 406 |
| National Institute of Diabetes and Digestive and Kidney Diseases | niddk.nih.gov | 193 |
| National Institute of Arthritis and Musculoskeletal and Skin Diseases | niams.nih.gov | 137 |
| National Heart, Lung, and Blood Institute | nhlbi.nih.gov | 84 |
| Eunice Kennedy Shriver National Institute of Child Health and Human Development | nichd.nih.gov | 78 |
| National Institute of Neurological Disorders and Stroke | ninds.nih.gov | 62 |
| Archived sites | archives.nih.gov | 38 |
| National Institute of Biomedical Imaging and Bioengineering | nibib.nih.gov | 21 |
| National Institute of Aging | nia.nih.gov | 17 |
| National Institute of Allergy and Infectious Diseases | niaid.nih.gov | 13 |
| Center for Information Technology | cit.nih.gov | 13 |
| National Institute of Dental and Craniofacial Research | nidcr.nih.gov | 13 |
| Fogarty International Center | fic.nih.gov | 6 |
| National Eye Institute | nei.nih.gov | 5 |
| National Institute on Deafness and Other Communication Disorders | nidcd.nih.gov | 5 |
| National Center for Advancing Translational Sciences | ncats.nih.gov | 4 |
| National Institute of Nursing Research | ninr.nih.gov | 3 |
| Center for Scientific Review | csr.nih.gov | 2 |
| NIH Clinical center | clinicalcenter.nih.gov | 1 |
| National Institute on Minority Health and Health Disparities | nimhd.nih.gov | 0 |
| National Center for Complementary and Integrative Health | nccih.nih.gov | 0 |

*** Figures obtained excluding Pubmed (-site:ncbi.nlm.nih.gov)**

b) U.S. Department of Health and Human Services (HHS) (http://www.hhs.gov)

Excluding citations, we find about 732 results by selecting all those documents indexed within the official website (site:hhs.gov). As with the NIH, we can find several subdomains: acf.hhs.gov (about 294 results), aspe.hhs.gov (238), oig.hhs.gov (56), minorityhealth.hhs.gov (19), ori.hhs.gov (16), archive.hhs.gov (17), ncvhs.hhs.gov (10), iacc.hhs.gov (5), and millionhearts.hhs.gov (2). A qualitative analysis of all these 732 documents (out of which 700 offer a link to the full text) would provide evidence about the coverage of gray literature within governmental agencies.

## 4. DISCUSSION AND CONCLUSIONS

The empirical results obtained in this study clearly differ from those obtained by Bonato (2016), and present a very different picture of the possibilities of GS as tool to find scientific information.

*Search features*

We entirely agree with Bonato on the issue of Google Scholar's limitations (fluctuation of results, lack of search filters, etc.). Obviously, Google Scholar lacks many of the advanced search features available in other bibliographic databases, such as Pubmed. These shortcomings have been already identified and described in the scientific literature, and recently compiled (Orduna-Malea et al, 2016). This makes search tedious and time-consuming tasks. However, it is one thing to have a friendly search experience, and quite another to find gray literature. Bonato claims that “Google Scholar also cannot be relied on to capture other gray literature search publications, such as guidelines published by governmental agencies”. The empirical results obtained in this work contradict this statement.

*Gray literature*

Gray literature - all those materials and research produced outside the commercial publishing and distribution channels - plays an important role in many fields of knowledge. For this reason, it was critical when this material was hard to find. The widespread use of the Web changed this, uncovering contents that were once considered “gray literature” but are now “public” (available online). However, contrary to the assertions by Bonato, gray literature never played such an important role in Biomedicine, with the exception of clinical trials, evidence-based documents, and consensus documents generated by the government and by scientific and professional associations.[2]

Moreover, conference papers and meeting abstracts published in conferences and other meetings are not essential information sources to the advancement of research and medical practice, because of the following reasons:

- A great percentage of the contents published in these events are eventually published in scientific journals.
- Unpublished documents are deposited in clinical trials databases, such as Clinicaltrials.gov (Thelwall and Kousha, 2016).
- Citation analyses have plainly proved that the impact of conferences is - in bibliometric terms - irrelevant, with exceptions in some areas, such as the computer sciences (Lisée, Lariviere and Archambault, 2008).

Extrapolation

Bonato uses three search strategies to solve the information need presented in the first case study (conference proceedings) and one to approach the second case study (guidelines). This seems insufficient to extrapolate conclusions to the entire database, especially when the advanced search options of the search engine were not taken into consideration. This leads the author to erroneous conclusions.

---

[2] We acknowledge a discrepancy on this assertion by one reviewer that must be worth noted. Below we show his/her blind suggestion:
- Contents may “eventually” be published in scientific journals, but not necessarily in a timely way.
- Only some investigators deposit unpublished results in ClinicalTrials.gov – and this only in recent years.
- Results that are not overwhelmingly positive may not be published in a journal article and may not be highly cited, but may still provide important evidence regarding the benefits/lack of benefit of a particular therapy.

Lastly, we conclude that Google Scholar is a powerful tool for searching gray literature, as long as the users are familiar with all the possibilities Google Scholar offers as a search engine. When searching on a database, especially one as popular and apparently easy to use as Google Scholar, it is important to make an adequate use of the available features; a poorly formulated search will undoubtedly return misleading results.

Bonato's work is considered just a subjective review and not a research paper under the editor's view. However, and beyond some methodological shortcomings highlighted in this working paper, its claims and statements are formally published in the academic system, spreading an erroneous conclusion about Google Scholar.

### Acknowledgment

The authors wish to formally express that we do not have any conflict of interest with Google Scholar, and our points of view are only determined by our objective empirical evidences while testing Google Scholar both as a source for scientific information discovery and a tool for scientific evaluation.